\documentclass[12pt,preprint]{aastex}

\slugcomment{}

\begin{document}

\title{Primordial $^4He$
Abundance Constrains the Possible Time Variation of the Higgs Vacuum Expectation Value.}


\author{Josef M. Ga{\ss}ner and Harald Lesch}
\affil{University Observatory Munich, Scheinerstr. 1, 81679 Munich, Germany}
\email{josef.gassner@skydsl.de}

\begin{abstract}
We constrain the possible time variation of the Higgs vacuum expectation value ($v$) by
recent results on the primordial $^4He$ abundance ($Y_P$). For that, we use an analytic
approach which enables us to take important issues into consideration, that have been
ignored by previous works, like the $v$-dependence of the relevant cross sections of
deuterium production and photodisintegration, including the full Klein- Nishina cross
section. Furthermore, we take a non-equilibrium Ansatz for the freeze-out concentration
of neutrons and protons and incorporate the latest results on the neutron decay. Finally,
we approximate the key-parameters of the primordial $^4He$ production (the mean lifetime
of the free neutron and the binding energy of the deuteron) by terms of $v\over v_0$
(where $v_0$ denotes the present theoretical estimate). Eventually, we derive the relation $Y_P \simeq 0.2479 - 5.54~ \left(\frac{v-v_0}{v_0}\right)^2 - 0.808~ \left(\frac{v-v_0}{v_0} \right)$ and the most stringent limit on a possible time variation of $v$ is given by: $-5.4 \cdot 10^{-4} ~ \leq~ \frac{v-v_0}{v_0} ~~ \leq 4.4 \cdot
10^{-4}$.
\end{abstract}



\section{Introduction}
The standard model \citep{Griffiths} is a remarkably successful description of
fundamental particle interactions.  The theory contains parameters - such as particle
masses - whose origins are still unknown and which cannot be predicted, but whose values
are constrained through their interactions with the so called Higgs field. The Higgs
field is assumed to have a non-zero value in the ground state of the universe - called
its vacuum expectation value $v$ - and elementary particles that interact with the Higgs
field obtain a mass
proportional to this fundamental constant of nature. \\
Although the question whether the fundamental constants are in fact constant, has a long
history of study (see \cite{Uzan} for a review), comparatively less interest \citep{Yoo,
Ichikawa, Kujat,
Scherrer,Dixit} has been directed towards a possible variation of $v$. \\
A macroscopic probe to determine the allowed variation range is given by the network of
nuclear interactions during the Big-Bang-Nucleosynthesis (see \cite{Eidelman_1} for a
review of the Standard Big-Bang-Nucleosynthesis Model), with its final abundance of
$^4He$. The relevant key-parameters are the freeze-out concentration of neutrons and
protons, the so called deuterium bottleneck (the effective start of the primordial
nucleosynthesis) and the neutron decay.\\
The major difference between our contribution and previous studies is, that we further
improve this key-parameters by using an analytic Ansatz, exclusively. This analytic
approach enables us to take important issues into consideration, that have been ignored
by previous works, like the $v$-dependence of the relevant cross sections of deuterium
production and photodisintegration, including the full Klein-Nishina cross section.
Furthermore, we take a non-equilibrium Ansatz for the freeze-out concentration of
neutrons and protons and incorporate the latest results on the neutron decay.

Finally we approximate the mean lifetime of the free neutron and the binding energy of
the deuteron by terms of $v$, to constrain its possible variation by recent results on
the primordial $^4He$ abundance
\citep{Fukugita,Olive_and_Skillman,Coc,Izotov,Izotov_Thuan,Luridiana}.

We briefly note, that constraints on the spacial variation of $v$ required a measurement
of helium abundance anisotropy or inhomogeneity versus the position in the sky and an
inhomogeneous theoretical BBN model. The homogeneous formalism used throughout the paper
thus assumes a spacial invariance of the Higgs vacuum expectation value.

\section{Calculations}

All relevant processes of SBBN took place at a very early epoch, when the energy density
was dominated by radiation, leading to a time-temperature relation for a flat universe:
\begin{eqnarray}\label{t_aus_T} t =
\sqrt{\frac{90 \hbar ^3 c^5}{32 \pi ^3 k^4 G g_*}} \frac{1}{T^2} ~[s],
\end{eqnarray}
where $c$ is the velocity of light, $k$ the Boltzmann constant, $G$ denotes the
gravitational constant and $\hbar$ is the Planck constant divided by 2$\pi$. $g_*$ counts
the total number of effectively massless ($mc^2 \ll kT$) degrees of freedom, given by
$g_* = \left( g_b + \frac{7}{8}g_f \right)$, in which $g_b$ represents the bosonic and
$g_f$ the fermionic contributions at the relevant temperature. The non-relativistic
species are neglected, since their energy density is exponentially smaller \citep{Kolb}.\\
\\
At very high temperatures ($T\gg 10^{10}$K), the neutrons and protons are kept in thermal
and chemical equilibrium by the weak interactions
\begin{eqnarray*}
(1)& n + e^+ &\rightleftarrows  p + \bar{\nu}_e, \\
(2)& n + \nu_e  &\rightleftarrows  p + e^- ~ { and }\\
(3)& n  &\rightleftarrows  p + e^- + \bar{\nu}_e,
\end{eqnarray*}
until the temperature drops to a certain level, at which the inverse reactions become
inefficient. This so called "freeze-out"-temperature $T_f$ and time $t_f$ denote the
start of the effective neutron beta decay and detailed calculations \citep{Mukhanov}
derive
\begin{eqnarray}\label{T_f}
\left( \frac{kT_f}{Q} \right)^2 \left( \frac{kT_f}{Q}+ 0.25 \right)^2 \simeq 0.18
\sqrt{\frac{\pi^2}{30}g_*(T_f)},
\end{eqnarray}
where Q is the energy-difference of the neutron and proton rest masses. At $T_f$ the
effectively massless species in the cosmic plasma are neutrinos, antineutrinos,
electrons, positrons and photons. For the case of three neutrino families $g_b=2$
and $g_f=10$, which gives $g_*(T_f)=10.75$.\\
\\
Following \cite{Mukhanov} we take a non-equilibrium freeze-out ratio of neutron number
density ($n_n$) to baryon number density ($n_N$):
\begin{eqnarray}\label{X_n}
\frac{n_n}{n_N}(T_f) = \int_0^{\infty} \frac{\exp\left[-5.42 (\frac{\pi^2}{30}
g_*(T_f))^{-\frac{1}{2}} \int_0^y (x + \frac{1}{4})^2(1+e^{-\frac{1}{x}})dx\right] }{2
y^2 (1+cosh(1/y))} dy,
\end{eqnarray}
where $y = k T_f / Q$. From now on the decay of free neutrons via $n \rightarrow p + e^-
+ \bar{\nu}_e$, with a mean lifetime \citep{Serebrov} $\tau_n=878.5$ sec, can no longer
be refreshed. Thus, whereas the neutron density decreases as $n_n(t) = n_n(t_f) \cdot
e^{-\frac{t-t_f}{\tau_n}}$, the proton density increases as $n_p(t) = n_p(t_f) +
(n_n(t_f) - n_n(t))$ and we obtain
\begin{eqnarray}\label{n_n/n_p}
\frac{n_p}{n_n}(t) &=& \frac{n_N}{n_n}(T_f) e^{\frac{t-t_f}{\tau_n}} - 1.
\end{eqnarray}
The next important step is the start of nucleosynthesis $t_N$, usually referred to as the
"deuterium bottleneck". The delay between $t_f$ and $t_N$ is caused by the very low
efficiency of direct production of light elements by successive collisions of several
free protons and neutrons to one nucleus. In fact, nucleosynthesis proceeds through
sequences of two-body reactions with the deuteron $d$ as the intermediate product, via $p
+ n \rightarrow d + \gamma$. Accordingly, the small binding energy of the deuteron
$B_d\simeq 2.225 $ MeV presents a severe problem for nucleosynthesis, since energetic
photons of the background radiation continuously disrupt the newly formed deuterons,
until the temperature drops to a certain level $T_N$, when the deuteron production gets
the upper hand over photodisintegration. Because the decaying neutrons can no longer be
refreshed by weak interactions after $t_f$, the interval between $t_f$ and $t_N$ plays an
essential role for the outcome of the primordial helium
production.\\
Hence, we have to calculate the rates of deuteron production $\Gamma_{(np\rightarrow
d\gamma)}$, deuteron photo-disintegration $\Gamma_{(d\gamma\rightarrow np)}$ and the
expansion rate of the universe $\Gamma_{exp}$, to determine $t_N$ respectively $T_N$,
when
\begin{eqnarray}\label{gammas}
\frac{\Gamma_{(np\rightarrow d\gamma)}}{\Gamma_{(d\gamma\rightarrow np)}} > 1 +
\frac{\Gamma_{exp}}{\Gamma_{(d\gamma\rightarrow np)}}.
\end{eqnarray}
$\Gamma_{exp}$ for a radiation-dominated, flat universe is given by $1 \over{2 t}$ with
t from Eq (\ref{t_aus_T}).\\
The rates for production and photo-disintegration of deuteron are given by the product of
the relevant number density, velocity and cross section ($\sigma$):
\begin{eqnarray}
\frac {\Gamma_{(np \rightarrow d\gamma)} }{\Gamma_{(d\gamma\rightarrow np)}} &=&
\frac{n_p~ \sqrt{\frac{3 k T}{m_N}}~ \sigma_{(np \rightarrow d\gamma)}}{n_{\gamma}^* ~c
~\sigma_{(d\gamma \rightarrow np)}} = \frac{\eta  ~ \sqrt{\frac{3 k T}{m_N}}~ \sigma_{(np
\rightarrow d\gamma)}}{\left(1+\frac{n_n}{n_p}(T)\right) \frac{n_{\gamma}^*}{n_\gamma} ~c
~\sigma_{(d\gamma \rightarrow np)}},
\end{eqnarray}
where $\eta\simeq 6.14 \cdot 10^{-10}$ is the baryon to photon ratio based on WMAP
\citep{Huey} and $n_{\gamma}^*$ denotes the number density of photons which effectively
disintegrate the deuteron. These photons have to supply enough energy and must not loose
this energy
in much more likely Compton scattering with electrons instead of deuterons.\\
The number density of photons at a certain temperature T is given by
\begin{eqnarray}
n_{\gamma}= \frac{8 \pi}{(h c)^3} \int_0^\infty \frac{E^2}{e^{\frac{E}{k T}}-1} dE = 16
\pi \zeta(3) \left(\frac{kT}{hc}\right)^3,
\end{eqnarray}
where $\zeta$ is the Riemann zeta function. The number density of these photons,
supplying a minimum energy $E_{dis} \gg kT$ is
\begin{eqnarray}
n_{(\gamma>E_{dis})} = \frac{8 \pi}{(h c)^3} \int_{E_{dis}}^\infty E^2 e^{-\frac{E}{k
T}}dE = 8 \pi \left(\frac{kT}{hc}\right)^3
\left[\left(\frac{E_{dis}}{kT}+1\right)^2+1\right] e^{-\frac{E_{dis}}{kT}}
\end{eqnarray}
whereas only a fraction $\frac{\sigma_{(d\gamma \rightarrow
np)}}{\frac{n_p}{n_n}(T)\sigma_{(e\gamma \rightarrow e\gamma)}}$ will successfully
disintegrate a deuteron leading to
\begin{eqnarray}
n_{\gamma}^* = n_{(\gamma>E_{dis})} \frac{\sigma_{(d\gamma \rightarrow
np)}}{\frac{n_p}{n_n}(T)\sigma_{(e\gamma \rightarrow e\gamma)}}.
\end{eqnarray}
We take $\sigma_{(e\gamma \rightarrow e\gamma)} \simeq \sigma_{KN}(E_{\gamma})$ where
$\sigma_{KN}$ denotes the Klein-Nishina cross section \citep{Rybicki} for electron photon
scattering and the mean incident photon energy $E_\gamma$ is given by
\begin{eqnarray}
E_\gamma &=& \frac{1}{n_{(\gamma>E_{dis})}}\frac{8 \pi}{(h c)^3} \int_{E_{dis}}^\infty
E^3 e^{-\frac{E}{k T}} dE
~\simeq~ E_{dis}+ k T. \label{E_gamma}
\end{eqnarray}

The interaction cross section of photo-disintegration $\sigma_{(d\gamma \rightarrow np)}$
can be derived by the calculations of \cite{Rustgi}. We use a least mean square
approximation within the incident photon energy range of 2.3 to 3.6 MeV and obtain
\begin{eqnarray}
\sigma_{(d\gamma \rightarrow np)} &=& \frac{1}{n_{(\gamma>E_{dis})}}\frac{8 \pi}{(h c)^3}
\int_{E_{dis}}^\infty E^2 e^{-\frac{E}{k T}} [-2162.3 + 8.1208 \cdot 10^{15} (E +
1.5 k T)] dE\\
\nonumber \\
&\simeq& -2162.3+ 8.1208 \cdot 10^{15} (E_{dis}+2.5 k T)~ [\mu b].
\end{eqnarray}
The cross section for neutron capture $\sigma_{(np \rightarrow d\gamma)}$ is related to
$\sigma_{(d\gamma \rightarrow np)}$ by \citep{Chen}:
\begin{eqnarray}
\frac{\sigma_{(np \rightarrow d\gamma)}}{\sigma_{(d\gamma \rightarrow np)}} \simeq
\frac{3 E_\gamma^2}{2 m_N c^2(E_\gamma - B_d)},
\end{eqnarray}
where we take $E_\gamma$ from Eq (\ref{E_gamma}). Collecting all terms, inserting into Eq
(\ref{gammas}) and taking into account, that $E_{dis}=B_d - \frac{3}{2} k T$ we obtain an
equation for $T_N$:
\begin{eqnarray}\label{T_N}
\left(\frac{ \eta \zeta(3)\frac{n_p}{n_n}(T_N)}{1+\frac{n_n}{n_p}(T_N)} \right)
\sqrt{\frac{(3kT_N)^3}{m_n^3 c^6}} \frac{\sigma_{(e\gamma\rightarrow e\gamma)}
e^{\frac{B_d}{kT_N}-\frac{3}{2}}}{\sigma_{(d\gamma\rightarrow np)}} =& \nonumber \\ = 1 +
\frac{n_p}{n_n}(T_N) \pi^2 & \sqrt{\frac{G g_*(T_N) h^3}{90
c}}\frac{\sigma_{(e\gamma\rightarrow e\gamma)}kT_N
e^{\frac{B_d}{kT_N}-\frac{3}{2}}}{\sigma_{(d\gamma\rightarrow np)}^2 10^{-34}
(B_d-\frac{1}{2}kT_N)^2}
\end{eqnarray}
where
\begin{eqnarray}
\frac{n_p}{n_n}(T_N) = \frac{n_N}{n_n}(T_f) \exp{\left[\frac{1}{\tau_n} \sqrt{\frac{90
\hbar ^3 c^5}{32 \pi ^3 k^4 G}} \left(\frac{1}{\sqrt{g_*(T_N)}T_N^2}-
\frac{1}{\sqrt{g_*(T_f)}T_f^2}\right)\right]}-1
\end{eqnarray}
and $\frac{n_N}{n_n}(T_f)$ is the reciprocal freeze-out ratio, given by Eq (\ref{X_n}).
The factor $10^{-34}$ on the right hand side of Eq (\ref{T_N}) is due to the unit $\mu b$
that we use for all cross sections $\sigma$.

The neutrinos have decoupled from equilibrium at about one MeV (above the rest mass
energy of an electron) and thus before the annihilation of electron positron pairs.
Therefore the entropy due to this annihilation is transferred exclusively to the photons,
i.e.
$g_*(T_N)\simeq3.36$.\\
\\
Once sufficient deuteron has been produced, all other reactions
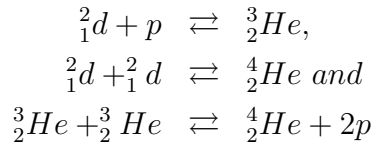
\begin{eqnarray*}
_1^2d + p & \rightleftarrows & _2^3He,\\
_1^2d + _1^2d & \rightleftarrows & _2^4He ~{ and }\\
_2^3He + _2^3He & \rightleftarrows & _2^4He + 2p
\end{eqnarray*}
proceed with significantly higher binding energies and the nucleosynthesis is no longer
constrained by photo-disintegration. It ends when the thermal energy is insufficient to
permit the energetically favored synthesis reaction - the fusion of deuteron.

With the assumption that at $t_N$ all available neutrons (as well as the same number of
protons) have been synthesized to $^4He$, which is not further transformed into heavier
nuclei, because elements with nucleon mass number A=5 to A=8 are insufficiently stable to
function successfully as intermediate products for nucleosynthesis at the available
densities, we can calculate $Y_P$, the final $^4He$ abundance by weight:
\begin{eqnarray}\label{Y_p}
Y_P &=&  \frac{2}{1+\frac{n_p}{n_n}(t_N)} = \frac{2 \frac{n_n}{n_N}(T_f)}{e^{\frac{t_N -
t_f}{\tau_n}}} = \frac{2 \frac{n_n}{n_N}(T_f)}{\exp{\left[\frac{1}{\tau_n}\sqrt{\frac{90
\hbar ^3 c^5}{32 \pi ^3 k^4 G}} \left(
\frac{1}{\sqrt{g_*(T_N)}T_N^2}-\frac{1}{\sqrt{g_*(T_f)}T_f^2} \right)\right]}},
\end{eqnarray}
where $\frac{n_n}{n_N}(T_f) \simeq 0.15709$ is given by Eq (\ref{X_n}), $T_f \simeq
\frac{Q}{k}~0.64794$ is given by Eq (\ref{T_f}) and Eq (\ref{T_N}) determines $T_N$,
respectively.

For comparison with the most recent numerical result \citep{Cuoco} $Y_P^{num}=0.2483$,
which assumes a mean neutron lifetime $\tau_n = 885.7 ~[s]$, we obtain (Eq \ref{Y_p}): $Y_P (\tau_n=885.7) = 0.2483$.\\
For comparison with the observation-based result \citep{Coc} $Y_P^{obs}=0.2479$, we take
\citep{Serebrov} $\tau_n = 878.5~[s]$ and obtain (Eq \ref{Y_p}): $Y_P
(\tau_n=878.5)=0.2479$. Furthermore, this calculation shows, how sensitive $Y_P$ depends
on the mean lifetime of neutrons.

Taking into account our simple approach for the start of nucleosynthesis (Eq
\ref{gammas}), where we neglected the fact, that the deuterons are not only destroyed by
photo-disintegration but also consumed by the fusion of light elements, the concordance
with the numerical as well as the observation-based result is very encouraging.

This analytic expression for $Y_P$ therefore provides our basis for finding the
dependence of $Y_P$ and the possible deviation of $v$ from its present value $v_0$, in
order to finally
constrain $v\over v_0$ by recent results on the primordial $^4He$ abundance.\\
\\
Crucial for the result of primordial nucleosynthesis is the moment $t_N$ or the
corresponding temperature $T_N$, at which the production rate gets the upper hand. $T_N$
depends on the binding energy of the deuteron $B_d$. Hence we take the linear fit of
$B_d$ versus $m_\pi$ that has been used by \cite{Yoo} and \cite{Mueller}, based on
\cite{Beane}:
\begin{eqnarray}
B_d(v) \simeq B_d(v_0)~(11-10 ~\frac{m_\pi}{m_{\pi_0}}),
\end{eqnarray}
where $m_\pi$ is the pion mass (the index 0 again denotes the present value). As
emphasized by \cite{Yoo}, in our narrow range of interest, $m_\pi^2 \propto v$, leading
to the final expression
\begin{eqnarray}
B_d(v) \simeq B_d(v_0)~(11-10 \sqrt{v/v_0}).
\end{eqnarray}

As $B_d$ changes, $E_{dis}$, $E_\gamma$ and the cross sections $\sigma_{(d\gamma
\rightarrow np)}$, $\sigma_{(np \rightarrow d\gamma)}$ and $\sigma_{(e\gamma \rightarrow
e\gamma)}$ change, accordingly. Concerning $\sigma_{(e\gamma \rightarrow e\gamma)}$ we
furthermore have to consider, that the mass of the electron varies proportionally
\begin{eqnarray}
m_e(v) = m_e(v_0)~\frac{v}{v_0}
\end{eqnarray}
which enters the Klein-Nishina cross section.

Any variation of $v$, of course, changes the value of Q as well, but according to Eq
(\ref{T_f}), $T_f$ is proportional to Q with the interesting consequence, that the
freeze-out concentration of neutrons and protons does not change with a varying Q.\\
By contrast, the Higgs vacuum expectation value definitely influences the mean lifetime
of the free neutrons $\tau_n$ and following \citep{Mueller} we use the expression
\begin{eqnarray}
\frac {\tau_n - \tau_{n_0}}{\tau_{n_0}} = 3.86~\frac{\alpha-\alpha_0}{\alpha_0}+~ 4
~\frac{v-v_0}{v_0}+ 1.52 ~\frac{m_e-m_{e_0}}{m_{e_0}} - 10.4
\frac{(m_d-m_u)-(m_{d_0}-m_{u_0})}{(m_{d_0}-m_{u_0})},
\end{eqnarray}
where $\alpha$ is the electromagnetic fine structure constant and $m_d$ and $m_u$ are the
masses of the up- and down-quark (the index 0 again denotes the present values). This
approximation is the result of a linear analysis, based on the assumption, that only one
single fundamental coupling changes with time while keeping the others fixed,
respectively. Furthermore, M\"{u}ller et al. only consider standard model particles (with
three neutrino
families) to contribute to the energy density at BBN.\\
Taking into account, that the elementary masses of the electrons and quarks linearly
depend on $v$ and disregarding the effect of a varying $\alpha$ (we take $\alpha$ as
constant throughout this letter\footnote{We also take $\hbar$, k, c, G and $\eta$ as
constant throughout the letter.}), we achieve
\begin{eqnarray}
\tau_n(v) \simeq \tau_n(v_0)~( 1 - 4.88 ~\frac{v-v_0}{v_0}).
\end{eqnarray}
 Finally, we derive a relation
between $Y_P$ and $v$, to constrain the permitted variation of the Higgs vacuum
expectation value by the primordial $^4He$ abundance:
\begin{eqnarray}
Y_P \simeq 0.2479 - 5.54~ \left(\frac{v-v_0}{v_0}\right)^2 - 0.808~ \left(\frac{v-v_0}{v_0} \right).
\end{eqnarray}
The predominant effect is the variation of the binding energy of the deuteron, followed by the mean neutron lifetime, whereas the changing mass of the electron with its consequences on the Klein-Nishina cross section (and the Compton scattering respectively) is almost negligible. The relative weights can be quantified by the linearisation
\begin{eqnarray}
Y_P \simeq 0.2479 + 0.105~ \left(\frac{B_d-B_{d_0}}{B_{d_0}}\right) + 0.058 ~ \left(\frac{\tau_n-\tau_{n_0}}{\tau_{n_0}} \right) - 0.006~\left(\frac{m_e-m_{e_0}}{m_{e_0}} \right).
\end{eqnarray}

\section{Results}
Using the observation based results of \citet{Coc} we derive
\begin{eqnarray*}
-5.4 \cdot 10^{-4} ~ &\leq&~  \frac{v-v_0}{v_0} ~~ \leq 4.4 \cdot 10^{-4}.
\end{eqnarray*}
We avoid the term "observational results" because all cited publications more or less
consist of interpretation of observational $^4He$-abundance plus theoretical input and
constraints by the cosmic microwave background. Especially the different interpretation
as a result of the deficiently understood systematics lead to incompatible data.
Therefore, we separately state all publications and their constraints on $v$ in the
following table:\\
\\
\begin{tabular}{|c|l|r|}
  \hline
  Authors & ~~~~~~~~$Y_P$   & Permitted variation $(v-v_0) \over v_0$\\
  \hline
  \cite{Fukugita} & $ 0.250~ \pm 0.004$              & $(-8.0^{~+5.0}_{~-5.3})\cdot 10^{-3}$\\
  \cite{Olive_and_Skillman} & $ 0.2491 \pm 0.0091$              & $(-1.5^{~+10.6}_{~-12.5})\cdot 10^{-3}$\\
  \cite{Coc}                    & $ 0.2479 \pm 0.0004$              & $(-0.49^{~+4.9}_{~-4.9})\cdot 10^{-4}$\\
  \cite{Izotov}              & $ 0.2443 \pm 0.0015$              & $(4.3^{~+1.7}_{~-1.7})\cdot 10^{-3}$\\
  \cite{Izotov_Thuan}         & $ 0.2421 \pm 0.0021$              & $(6.8^{~+2.3}_{~-2.3})\cdot 10^{-3}$\\
  \cite{Luridiana}        & $ 0.2391 \pm 0.0020$              & $(10.1^{~+2.1}_{~-2.2})\cdot 10^{-3}$\\

  \hline
\end{tabular}\\
\\
\\
Taking $v_0$=$v$ as constant, the primordial $^4He$ abundance can be used for another
interesting subject. The Higgs vacuum expectation value can be calculated theoretically
by \citep{Dixit}
\begin{eqnarray}
v_0 = 2^{-1/4} G_f^{-1/2} \simeq 246.22~ [GeV],
\end{eqnarray}
where $G_f \simeq 1.166371 ~[GeV^{-2}]$ is the Fermi coupling constant
\citep{Eidelman_2}, based on measurements of the muon mass and lifetime. The uncertainty
of determining
$v_0$, especially the contributions of higher order terms, can now be constrained:\\
\\
\begin{tabular}{|c|c|}
  \hline
  Authors & Compatible $v_0$ in GeV\\
  \hline
  \cite{Fukugita} & $ 245.56^{~+ 1.22}_{~-1.31}$\\
  \cite{Olive_and_Skillman} & $ 245.84^{~+ 2.63}_{~-3.08}$\\
  \cite{Coc}                    & $ 246.21^{~+ 0.12}_{~-0.12}$\\
  \cite{Izotov}              & $ 247.28^{~+ 0.42}_{~-0.44}$\\
  \cite{Izotov_Thuan}         & $ 247.90^{~+ 0.57}_{~-0.59}$\\
  \cite{Luridiana}        & $ 248.72^{~+ 0.52}_{~-0.54}$\\
  \hline
\end{tabular}\\
\\


\acknowledgments

We thank K. Butler and A. Jessner for valuable comments on the manuscript.






\appendix




\clearpage



\end{document}